\documentclass[twoside,leqno,twocolumn]{article}

\usepackage[letterpaper]{geometry}
\usepackage{multirow}
\usepackage{longtable}
\usepackage{ltexpprt}
\usepackage{amsmath,amssymb,mathtools}
\usepackage{algorithm}
\usepackage[noend]{algpseudocode}
\usepackage{multirow,array}

\providecommand{\keywords}[1]
{
\textbf{Keywords ---} #1
}

\begin{document}

\title{\Large A Game Theoretic Algorithm for Elite Customer Identification in Online Fashion E-Commerce
}
\author{
Chandramouli K\thanks{Myntra Designs Pvt. Ltd.}
\and Gopinath A\thanks{Myntra Designs Pvt. Ltd.}
\and Girish Satyanarayana \thanks{Myntra Designs Pvt. Ltd.}
\and Ravindra babu Tallamraju \thanks{Myntra Designs Pvt. Ltd.}
}

\date{}

\maketitle


\fancyfoot[R]{\scriptsize{Copyright \textcopyright\ 2021 by SIAM\\
Unauthorized reproduction of this article is prohibited}}





\begin{abstract} \small\baselineskip=9pt 
Myntra is an online fashion e-commerce company based in India. At Myntra, a market leader in fashion e-commerce in India, customer experience is paramount and a significant portion of our resources are dedicated to it. Here we describe an algorithm that identifies eligible customers to enable preferential product return processing for them by Myntra. We declare the group of aforementioned eligible customers on the platform as elite customers. Our algorithm to identify eligible/elite customers is based on sound principles of game theory. It is simple, easy to implement and scalable.
\end{abstract}

\vspace{0.05cm}

\section{Introduction}
Game theory is a method of studying strategic interactions. It has it's origins in the works of John von Neumann
and Oskar Morgenstern \cite{OandJ}. Later works by prominent mathematicians and economists like John Nash, Reinhard Selten, Eric Maskin, Roger Myerson  etc., established the field of game theory with applications to a diverse range of areas like computer science, social science, biology, logic etc. 

Scientific study of customer interaction is an important problem in e-commerce. For e.g.  \cite{Ge} identifies factors that play a role in customer loyalty towards e-commerce. \cite{Ke} focuses on the value of e-commerce to a customer. \cite{HK} describes interface design for better customer interaction with the e-commerce platform. \cite{BG} empirically investigates the impact of social media on customer engagement in e-commerce.  

One way to mathematically model and investigate customer interaction with an e-commerce platform is via game theory. For e.g.
\cite{RDV} analyses customer behaviour as a function of reputation of the seller in e-commerce systems utilizing game theory methodologies.
\cite{GM} proposes tools based on game theory for better customer satisfaction in e-commerce. \cite{LLZ} models customer's after service interaction as a game and analyses it and suggests promotion strategies.

Here in our work in fashion e-commerce, we apply equilibrium concepts developed by Nash and Selten and design a novel classification algorithm for elite customer identification. In particular we model the customer-Myntra relationship as a repeated game and identify equilibrium strategy vectors. Based on this analysis, we then design a classification algorithm that identifies eligible/elite customers to enable preferential return processing for them. Similar type of work is present in economics literature. For e.g.\cite{CCB} models interaction between two anonymous economies as a repeated game and analyses equilibrium strategies. However our novelty is in the application of similar techniques in fashion e-commerce and in particular modeling customer-Myntra interaction. Our key contributions are summarized below. 

\subsection*{Our contributions:}
\begin{itemize}
    \item We model the problem of elite customer identification at Myntra as a repeated game.
    \item We solve the game for an equilibrium strategy.
    \item With the equilibrium strategy as a guide we design an algorithm that identifies elite customers.
    \item We compare and evaluate the algorithm on real world data available at Myntra.
\end{itemize}

\section{Background}

In this section we present essential background \cite{YN} for our solution of the problem discussed.
To start a strategic form game $\Gamma$ is a tuple $(N, (A_i)_{i \in N}, (u_i)_{i \in N})$. Here 
\begin{itemize}
    \item $N=\{1,2,\cdots,n\}$ denotes a finite set of players. 
    \item $A_{1},A_{2},\cdots,A_{n}$ are the strategy/action sets of the players. 
    \item $u_{i}:A_{1} \times A_{2}\times \cdots A_{n} \rightarrow \mathbb{R}, \forall i \in N$, denote utility functions of the players. 
\end{itemize}
Let $A \coloneqq A_{1} \times A_{2}\times \cdots A_{n}$ denote the set of strategy vectors. A typical strategy vector is $(a_{1},a_{2},\cdots,a_{n})$ where $a_{i}$ denotes the strategy of player $i.$ Also denote by $A_{-i}$, the Cartesian product $A_{1} \times \cdots \times A_{i-1} \times A_{i+1} \times \cdots A_{n}$. Therefore $A_{-i}$ is the set of strategy vectors that consists of strategies of all players other than $i$ and a typical strategy vector is of the form $(a_{1},a_{2},\cdots,a_{i-1},a_{i+1},\cdots,a_{n})$. Moreover $(a_{i},a_{-i})$ is a complete strategy vector for the game.
\par
Given a strategic form game $\Gamma=(N, (A_i)_{i \in N}, (u_i)_{i \in N})$ and a strategy vector $a_{-i} \in A_{-i}$, we call $a_{i} \in \arg\max_{a'_{i} \in A_{i}} u_{i}(a'_{i},s_{-i})$, a best response strategy of player $i$ given $a_{-i}.$ In simple terms, $a_{i}$ is a strategy that maximizes player $i$'s utility given the strategies of all other players in the game.
\par
Given a strategic form game $\Gamma=(N, (A_i)_{i \in N}, (u_i)_{i \in N})$, a strategy vector $a^{*}=(a^{*}_{1},a^{*}_{2},\cdots,a^{*}_{n})$ is called a pure strategy Nash equilibrium of $\Gamma$ if 
$$u_{i}(a^{*}_{i})=\max_{a_{i}\in A_{i}}u_{i}(a_{i},a^{*}_{-i}) ~ \forall i \in N.$$
In simple terms, in a strategy vector that is a Nash equilibrium every player has best response strategy given strategy vector of all other players.

It is important to note that Nash equilibrium is self-enforcing i.e., on the condition that every player other than player $i$ chooses his/her strategy in Nash equilibrium then player $i$ is better off with the strategy in Nash equilibrium. In other words Nash equilibrium secures cooperation among players.

In summary unilateral deviations from Nash equilibrium ensure detrimental payoff for the deviate. In a two-player game since there are only unilateral deviations of Nash equilibrium it is ideal that the second player play Nash equilibrium strategy on the condition that the first player plays Nash equilibrium strategy.

\section{Problem Statement}
At Myntra, a customer's product return request is processed in the following manner. Once a customer places a product return request, a doorstep pickup agent arrives and receives the product. After a quality check of the returned item at a designated location refund to the customer is initiated.

For a better customer experience we wish to identify elite customers and do away with this quality check process for these elite customers. To identify these elite customers, we model the interaction between Myntra and a generic customer as a game.

We have two players in our game i.e. $N=\{ \text{Myntra}, \text{customer}\}$. Myntra has two actions i.e. $A_{\text{Myntra}}=\{\text{immediate refund},\text{no immediate refund}\}$. Similarly customer has two actions i.e. $A_{\text{customer}}=\{\text{comply with return requirements}, \\ \text{don't comply with return requirements}\}$.
Here in the context of Myntra the strategy/action ``immediate refund" refers to the removal of quality check process.

We designed the utility matrix  given in Table \ref{utility_matrix} for this game.
Our rationality behind this particular choice for the utility matrix is as follows. For Myntra, with $u_{\text{Myntra}}\text{(No Immediate Refund, Don't Comply)}=0$ as baseline, $u_{\text{Myntra}}\text{(Immediate Refund, Comply)}=1$, 
$u_{\text{Myntra}}\text{(No Immediate Refund, Comply)}=2$ and $u_{\text{Myntra}}\text{(Immediate Refund, Don't Comply)}=-1$ reflect the satisfaction levels of Myntra in the customer-Myntra relationship. We assume the same for the customer and obtain the utility matrix shown in Table \ref{utility_matrix}.
  \begin{table}
    \setlength{\extrarowheight}{2pt}
    \begin{tabular}{*{4}{c|}}
      \multicolumn{2}{c}{} & \multicolumn{2}{c}{Customer}\\\cline{3-4}
      \multicolumn{1}{c}{} &  & Comply  & Don't Comply \\\cline{2-4}
      \multirow{2}*{Myntra}  & Immediate &    &  \\
      & Refund & $(1,1)$ & $(-1,2)$ \\\cline{2-4}
      & No Immediate &      &  \\
      & Refund & $(2,-1)$ & $(0,0)$ \\\cline{2-4}
    \end{tabular}
     \caption{Utility matrix of our game}
      \label{utility_matrix}
  \end{table}
 \vspace*{-\baselineskip}
 \section{Solution}
 Observe that (No Immediate Refund, Don't Comply) is a pure strategy Nash equilibrium for this game and No Immediate Refund is the corresponding strategy for Myntra. However we note that the relation between Myntra and the customer is an ongoing relation. Hence a repeated game that repeats with probability $\delta$ is a more appropriate model for Myntra-customer relationship.
 In this context the set of strategy vectors is given by $S= (A_{\text{Myntra}} \times A_{\text{customer}})^{\mathbb{N}}$, set of sequences with values in $A_{\text{Myntra}} \times A_{\text{customer}}.$ Here $\mathbb{N}$ denotes the set of Natural numbers. The utility function $v_{\text{Myntra}}:S\rightarrow \mathbb{R}$ is given by 
 $v_{\text{Myntra}}(s)=\sum_{k\geq0}\delta^{k} u_{\text{Myntra}}(s^{k+1})$ where $s=(s^{1},s^{2},\cdots)$ with $s^{k} \in A_{\text{Myntra}} \times A_{\text{customer}}, k \in \mathbb{N}$ and $u_{\text{Myntra}}$ is given by Table \ref{utility_matrix}. Similarly the utility function 
 $v_{\text{customer}}:S\rightarrow \mathbb{R}$ is given by 
 $v_{\text{customer}}(s)=\sum_{k\geq0}\delta^{k} u_{\text{customer}}(s^{k})$ where $s=(s^{1},s^{2},\cdots)$ with $s^{k} \in A_{\text{Myntra}} \times A_{\text{customer}}, k \in \mathbb{N}$ and $u_{\text{customer}}$ is given by Table \ref{utility_matrix}.
 
 Observe that in this setup the strategy vector $s^{*}=(s^{*i})_{i\in \mathbb{N}}$ with $s^{*i}=(\text{no immediate refund},\text{don't comply})$ is an equilibrium strategy vector with $(v_{\text{Myntra}},v_{\text{customer}})=(0,0)$ as any unilateral deviation in any repetition of the game by a player diminishes the utility of the player. Moreover the equilibrium strategy ``no immediate refund" is currently employed by Myntra.
 This strategy however is customer independent and treats all customers alike and does not enable identification of elite customers as well as preferential returns processing for elite customers.

 Consider the following strategy $s^*=(s^{*1},s^{*2},\cdots)$ with $s^{*1}$=(immediate refund, comply) and $s^{*k}=s^{*(k-1)}$ if $s^{*(k-1)}$=(immediate refund, comply) else (no immediate refund, don't comply), that is cooperate to start with and don't cooperate once non-cooperation is observed. We note that for this strategy the utility obtained is given by
 \begin{align*}
  &(v_{\text{Myntra}},v_{\text{customer}}) \\
 =&\left(\displaystyle\sum_{k\geq0}\delta^{k} u_{\text{Myntra}}(s^{*k}),\displaystyle\sum_{k\geq0}\delta^{k} u_{\text{customer}}(s^{*k})\right) \\
 =&\left(\displaystyle\sum_{k\geq0}\delta^{k}1^k,\displaystyle\sum_{k\geq0}\delta^{k}1^k\right)
 =\left(\frac{1}{1-\delta},\frac{1}{1-\delta}\right)
 \end{align*}
 Now for any unilateral deviation by a player, at any game repetition stage $k$, the utility $v_{i}=\sum_{0\leq l\leq k-1}\delta^{l}1^l+2\delta^{k}, i\in N$. For this $s^*$ to be an equilibrium, from the definition, we require 
 \begin{align*}
    \displaystyle\sum_{0\leq l\leq k-1}\delta^{l}1^l+2\delta^{k} &\leq \frac{1}{1-\delta} \\
    \implies \frac{1-\delta^{k}}{1-\delta}+2\delta^{k} &\leq \frac{1}{1-\delta}
    \implies  \delta \geq \frac{1}{2}
 \end{align*}
 Hence our strategy $s^*$ is a pure strategy Nash equilibrium if the probability of repetition of the game is at least $0.5$.
 
 On a separate note, a relevant equilibrium concept for the case of repeated games is subgame perfect Nash equilibrium \cite{YN}. We also note that $s^*$ can be shown to be a subgame perfect Nash equilibrium. 
 
 We note here an important observation. Our model of Myntra-customer relationship as a repeated game apart from explaining current Myntra strategy as an equilibrium strategy-there by validating the choice of utlity matrix- suggests an alternative equilibrium strategy vector given by $s^*$. 
 
 We also note here that there are other equilibrium strategies as well. For e.g.
 $s=(s^1,s^2,\cdots)$ with $s^1=$(immediate refund, comply) and $s^{k}=$(immediate refund, comply) if $s^{k-1}=$(immediate refund, comply) or (no immediate refund, does not comply) else $s^{k}=$(no immediate refund, don't comply) is an equilibrium strategy provided $\delta=1$ and each such equilibrium strategy gives an algorithm to identify elite customers. Here we chose a strategy that is most cautious from the point of view of Myntra.

 With this analysis in place we design the following algorithm that enables us to identify elite customers
 
\section{Algorithm}
\label{algorithm}
\begin{algorithm}[ht]
\caption{Identify Elite Customers}
\begin{flushleft}
\textbf{Input:}\\ 
\text{Historic return request compliance status of the customers}\\ 
\text{and purchase sequence of the customers on Myntra} \\  
\text{Threshold $\tau=0.5$}\\
\textbf{Output:} \\ 
\text{Eliteness of the customer}
\end{flushleft}
\begin{algorithmic}[1]
\Procedure{CLASSIFY CUSTOMERS:}{}
\For{each customer}
\State Etimate the probability of repetition $\delta$ of the game from historic purchase sequence data.
\If{$\delta <\tau$} 
{customer is not elite}
\Else\If{customer complied in all returns}
    {customer is elite}
    \Else
        { customer is not elite}
    \EndIf
\EndIf
\EndFor
\State \textbf{return} Eliteness for customers 
\EndProcedure
\end{algorithmic}
\label{alg:Classify}
\end{algorithm}
Our algorithm given the sequence of purchases of a customer in 9 months into the past on the  platform estimates the probability of repetition $\delta$. We say that the game is repeated on the condition that the time gap between consecutive purchases of the customer is less than or equal to 90 days. We count the number of times the game is repeated utilizing the purchase sequence of the customer and estimate the probability of repetition $\delta$. The choices -9 months, 90 days- for time periods to estimate $\delta$ are motivated by business requirements.

We collect the historical product return compliance status of the customer. With these quantities at its disposal the algorithm compares the repetition probability estimate with the threshold $\tau$ and assigns eliteness to the customer. We note that with our utility matrix we get that $\tau = 0.5$.

We note that our algorithm is very intuitive. In summary it classifies a customer as elite if his frequency of purchases is more than 0.5 and has perfect returns compliance history. We also note that our algorithm is dynamic. As the 9 month window is dynamic the elite group of customers changes with time and a given customer is required to be a frequent purchaser from Myntra to be consistently elite.

\section{Experiments}
We evaluated our algorithm on historical real world e-commerce data available at Myntra. We note two types of error in our classification. We classify a customer as elite and the customer doesn't comply, we call them as false-positives (fp). We call as false-negatives (fn) all those customers that are not elite but comply. Similarly we define true-positives (tp) and true-negatives (tn) in an analogous manner. With these definitions in place we compute precision and recall and evaluate our algorithm

We choose the day October 2, 2020 to evaluate our algorithm. We obtained for each customer that placed a return request on the Myntra platform on this day the repetition probability estimate and also product return compliance status of the customer in the past 9 months. We classified the customer according to Algorithm \ref{alg:Classify}

Our emphasis is on false-positives as it impacts returned product resale value. Hence we aim to maximize precision with a satisfactory recall. We note here that precision is given by $\frac{tp}{tp+fp}$ and recall is given by $\frac{tp}{tp+fn}$.

We compared our algorithm against classification algorithms like logistic regression and random forests. For these algorithms some of the important features are summarised in Table \ref{features}. We note that our algorithm has better recall compared to logistic regression and random forests and all algorithms report close to perfect precision. We summarise the comparison in Table \ref{exp:analysis}

We note that our algorithm is similar to a decision tree. Here, unlike splitting of a node in a classical decision tree, we split the node based on equilibrium strategy given by the game. Hence it may be possible to fine tune hyper parameters of ensemble classification algorithms, for e.g., of random forests to achieve better performance. However unlike these classification algorithms our algorithm is easily explainable- it is easy to see the conditions under which a customer is elite, a desirable feature for businesses in the context of customer experience. In summary, the classification algorithm based on game theory methodologies achieves desired precision with a satisfactory recall and has the additional advantage of explainability.

\begin{table}[ht]
\centering
\begin{tabular}{|l|l|l|l|}
\hline
          & Our  & Logistic & Random \\
          & Algorithm & Regression & Forests \\ \hline
Precision & \textbf{99.7\%}              &     99.7\%                &    99.6\%        \\ \hline
Recall    & \textbf{59.0\%}            &     48.4\%            &   57.8\%            \\ \hline
\end{tabular}
\caption{Comparison of our classification algorithm}
\label{exp:analysis}
\end{table}

\begin{table}[ht]
\centering
\begin{tabular}{|l|l|}
\hline
\textbf{Features}                & \textbf{Description}                                                                 \\ \hline
\#Q2                         & No. of compliance failures \\
                            & in last 9 months by customer                                                                       \\ \hline
\#Q1                       & No. of compliance successes \\
                            & in last 9 months  by customer                                               \\ \hline
nserves          & No. of times the \\ 
                    & returned product is served by Myntra                                       \\ \hline
$\delta$                   & Estimated probability of \\
                            & repetition of the game                                                           \\ \hline
l\_status             & Compliance state of \\ 
                        & last return request by customer                                          \\ \hline
\end{tabular}
\caption{Important features of the classification algorithms}
\label{features}
\end{table}

\section{Conclusion}
We modeled the problem of identification of elite customers and preferential returns processing for them as a two-player repeated game and solved for its equilibrium strategy. We designed an algorithm based on this analysis and evaluated the algorithm on real-world data available at Myntra. We note that our algorithm is scalable, easy to implement and has performance comparable to classification algorithms like logistic regression and random forests. Moreover it has the advantage of explainability.

\bibliographystyle{plain}
\bibliography{references}
\end{document}